\documentclass[fleqn,twoside]{article}
\usepackage{espcrc2}

\usepackage{graphicx}
\usepackage{epsfig}
\usepackage[figuresright]{rotating}

\def \litwo {{\rm{Li_2}}}
\def \litr {{\rm{Li_3}}}
\def \lifo {{\rm{Li_4}}}

\newcommand{\AmS}{{\protect\the\textfont2
  A\kern-.1667em\lower.5ex\hbox{M}\kern-.125emS}}

\title{
{\normalsize \tt
DESY 05-151
\\
WUE-ITP-2005-009
\\
SFB/CPP-05-40
\\ August 2005}
\\
\vspace*{1.cm}
Harmonic polylogarithms for massive Bhabha scattering}
\author{M. Czakon%
\address[Wurzb]{Institut f\"ur Theoretische Physik
und Astrophysik, Universit\"at W\"urzburg,
Am Hubland, D-97074 W\"urzburg, Germany}
\address[Katowice]{Institute of Physics, Univ. of
    Silesia, Universytecka 4, 40007 Katowice, Poland},
J. Gluza%
\addressmark[Katowice]
\address[DESYZ]{DESY, Platanenallee 6, 15738
          Zeuthen, Germany},    
T. Riemann\addressmark[DESYZ]
}

\begin{document}

\begin{abstract}
One- and two-dimensional harmonic polylogarithms, HPLs and GPLs, appear
in calculations  of multi-loop integrals. We discuss them
in the context of analytical solutions for two-loop master integrals in
the case of massive Bhabha scattering in QED.
For the GPLs we discuss analytical representations,
conformal transformations, and also their transformations
corresponding to relations between master integrals in the
$s$- and $t$-channel.
\vspace{1pc}
\end{abstract}

\maketitle

\section{Introduction}

Multiloop calculations may be treated with numerical or analytical approaches.
The latter is limited to issues with few, typically at most three
different scales. 
But it has well-known attractive features, notably the knowledge
of the explicit structure of the singularities and a good control of
numerical stability.

The analytical approach to be discussed here relies on differential equations \cite{Kotikov:1991hm,Remiddi:1997ny}.
Iterative solutions in powers of $\epsilon = (4-D)/2$ are inherently connected with certain classes of special functions.
For two-scale problems (e. g. QED self-energies and vertices), the harmonic polylogarithms $H(\{a\},x)$ (HPLs) have been introduced
\cite{Remiddi:1999ew};
the index vector $\{a\}$ has elements $1,0,-1$.
We just mention that
HPLs with some generalized arguments are introduced in \cite{Aglietti:2003yc}.
A generalization of HPLs for three-scale problems (e.g. QED boxes) are the two-dimensional harmonic polylogarithms $G(\{b\},x)$ (GPLs);
the index vector $\{b\}$ has elements $1,0,-1$, but now also those depending on a second kinematic variable $y$.
This was observed in \cite{Gehrmann:1999as}
and systematically worked out in \cite{Gehrmann:2000zt}
 for a planar massless problem.
There one may find analytical expressions for the
GPLs until weight 3 with indexes $0, 1, z, 1-z$. Another generalization
is performed in \cite{Birthwright:2004kk}. Algebraic non-linear factors are introduced
as being motivated by the physical nature of the problem treated there.

Let us shortly review some facts on HPLs.
HPLs up to weight 4 are expressed, with few exceptions,
in terms of Nielsen Polylogarithms
 \cite{Remiddi:1999ew,Moch:1999eb,Czakon:2004wm}.
The exceptions are of weight 4 and may be expressed by few integrals \cite{Czakon:2004wm}:
\begin{eqnarray}
I_1(x)&=&
\int_0^x \frac{dy}{1+y} {\litr}(y),
\\
  I_2(x) &=& \int_0^x \frac{dy}{1+y} \litwo(y) \ln(1-y),
\\
  I_3(x)&=& \int_0^x \frac{dy}{1+y} \ln(y)\ln^2(1-y),
\\
  I_4(x) &=& \int_0^x \frac{dy}{1+y} \ln^2 (y) \ln(1-y),
\end{eqnarray}
where $I_1(x)=H[-1,0,0,1,x]$.
However, it cannot be excluded that some relations among them exist.
In fact, a relation between $I_3$ and
$I_4$ exists
\cite{Davydychev:2005nn}.
In fact, exploiting that  $\int dy \ln^3(y/(1-y))$ is known to Mathematica (after substituting  $y/(1-y)\to z$) and is related
to $I_3 - I_4$, one may derive
\begin{eqnarray}
&&I_4(x)= I_3(x)+\frac{1}{3} \Biggl[ \ln^3(1-x) \ln \left( \frac{1+x}{2} \right)
\nonumber \\
&-& \ln^3(x) \ln(1+x) + \ln^3 \left( \frac{x}{1-x} \right) \ln(1+x)
\Biggr]
\nonumber \\
&+& \ln^2 (1-x) \litwo \left( \frac{1-x}{2} \right)- \ln^2(x) \litwo (-x)
\nonumber \\
&-& 2\ln(1-x) \litr \left( \frac{1-x}{2} \right) +2 \ln(x) \litr(-x)
\nonumber \\
&+& \ln \left( \frac{x}{1-x} \right) \left[ \ln \left( \frac{x}{1-x} \right)
\right. \nonumber \\
&\times& \left( \litwo \left( \frac{2 x}{x-1} \right)-
\litwo \left( \frac{x}{x-1} \right) \right) \nonumber \\
&+&2 \left. \left( \litr \left( \frac{x}{x-1} \right)-
\litr \left( \frac{2 x}{x-1} \right) \right) \right] \nonumber \\
&+& 2 \left[ \lifo \left( \frac{1-x}{2} \right)- \lifo(1/2)-
\lifo (-x) \right. \nonumber \\
&-& \left. \lifo \left( \frac{x}{x-1} \right) +
\lifo \left( \frac{2 x}{x-1} \right) \right] .
\end{eqnarray}
As far as numerical solutions are concerned, the problem is solved for any
weight of HPLs. The numerical evaluation of HPLs with indexes
$\{0,1,-1\}$ (and the analytic continuation) 
is described in \cite{Gehrmann:2001pz}
and may be performed with the Fortran program {\tt hplog}. These HPLs
have been programmed also in Mathematica \cite{Maitre:2005uu}.
The systematic numerical treatment
of GPLs with indexes $\{0,1,1-z,-z\}$ 
is given in  \cite{Gehrmann:2001jv} and performed with
Fortran program {\tt tdhpl}.\footnote{This is in fact not
sufficient to cover our physical cases; we need indexes
$\{0,1,-1,1-z,-z\}$).} Please notice a change of notation compared to 
\cite{Gehrmann:2000zt} as described in the Appendix of \cite{Gehrmann:2001jv}.

In this paper, we will discuss features of the GPLs which we are exploiting for a study of massive QED two-loop boxes.
These GPLs were used in e.g.
\cite{Bonciani:2003te,Bonciani:2003cj} and
\cite{Czakon:2004wm,Czakon:2004tg,Czakon:2004wu}.
In the next section we describe the general idea of solving differential equations for Feynman integrals with GPLs, then we sketch a simple one loop
case, and what finally follows is a section on some algebraic relations connected with GPLs as we are using them for Bhabha scattering.
\section{Differential equations
}
Master integrals (MIs) $M$ may be determined as solutions of appropriate differential equations.
The general idea is simple.\footnote{A nice pedagogical introduction
  is \cite{Aglietti:2004vs}.}
Let $M$ fulfill some differential equation with respect to $x$,
\begin{equation}
\frac{d}{dx} M(x) = A(x) M(x) + B(x),
\end{equation}
where $x$ is
usually the external scale like $s$ (or $t$) or a conformal counterpart $x$ (or $y$); for definitions and details see \cite{Czakon:2004wm}.
The inhomogeneity $B$ is a linear combination of MIs of lower complexity and assumed to be known from an earlier stage of iteration.
For simplicity we assume here $M$ to depend only on $x$.
If $H(x)$ is a solution of the homogeneous equation,
\begin{equation}
\frac{d}{dx} H(x) = A(x) H(x),
\label{heq}
\end{equation}
then the full solution is given by
\begin{equation}
M(x) = H(x) \left[ Const + \int^x  \frac{dx'}{H(x')} B(x') \right] .
\label{deq}
\end{equation}
The solutions in arbitrary dimension $D$ are generally combinations
of generalized hypergeometric functions
which are difficult (or at least tedious) to find and to expand in powers of $\epsilon$.
This statement is true already
for the massive QED one-loop box \cite{Fleischer:2003bg}, where
Appell hypergeometric functions and a Kamp\'e de F\'eriet function
appear.
An alternative idea is to expand (\ref{deq}) in $\epsilon$:
\begin{eqnarray}
M&=&\sum_{i=-\alpha}^{\beta}{m^{i} \epsilon^{i}}, \\
\label{ai}
A&=&\sum_{i=0}^{\alpha+\beta}{a^{i} \epsilon^{i}} ,\;\;\;
B=\sum_{i=-\alpha}^{\beta}{b^{i} \epsilon^{i}} ,
\end{eqnarray}
where $\alpha$ is fixed by the physical problem and $\beta$ chosen reasonably.
Then one may try to solve iteratively the system of equations for the $m^i(x)$:
\begin{equation}
\label{mi}
\frac{d}{dx} m^i (x) = \sum_{j=0}^{\alpha-i} a^{j}(x) ~m^{i-j}(x) + b^{i}(x).
\end{equation}
Let us mention that $A$ should not be singular in $\epsilon$ ($i \geq 0$ in (\ref{ai})).
Otherwise some $m^i(x)$ on the LHS of (\ref{mi})
would depend on a higher order component $m^{i-j}(x)$ of $M$ sitting
on the RHS, and the recursion could not be solved. With this assumption the solution is of the form
\begin{eqnarray}
m^i (x) &=& H(x) \Biggl( Const +
\\
&& \hspace{-1.5cm} \left. \int^x \frac{dx'}{H(x')} \left[
\sum_{j=1}^{\alpha-i} a^{j}(x) m^{i-j}(x) + b^{i}(x) \right]
\right)  .
\nonumber
\end{eqnarray}
$H(x)$ is the solution of the homogeneous equation (\ref{heq}) and is the same
for all orders in $\epsilon$.

In general, some sets of
MIs will fulfill a system of linear differential
equations.
For massive Bhabha scattering so far only one such system {\tt B5l4m}, the
two-loop boxes with 5 lines, 4 of them massive, with two MIs has been fully solved analytically in the language
of HPls and GPLs \cite{Czakon:2004wm,Bonciani:2003cj}.
Another analytically fully solved 4-point MI is {\tt B5l2m1}
\cite{Czakon:2004tg}.
We will comment on  the more complex  system of 4 MIs {\tt B5l2m3}
in the context of GPLs in Section \ref{secgpls}.
\section{The Bhabha one-loop box: An illustration}
A differential equation for the one-loop box {\tt B4l2m}
is \footnote{For notations see \cite{Czakon:2004wm}.}
\begin{eqnarray}
&&s \frac{\partial}{\partial s} {\tt B4l2m[s,t]}=a\; {\tt B4l2m[s,t]}+b\;
{\tt SE2l2m[s]} \nonumber  \\
&& +
c\; {\tt T1l1m} + d \; {\tt SE2l0m[t]} +e \; {\tt V3l1m[s]}.
\label{box}
\end{eqnarray}
The tadpole {\tt T}, self energies {\tt SE}, and vertex {\tt V} are known.
The factor $a$ of the homogeneous part is
\begin{eqnarray*}
a&=& \frac{8 + s^2 - 2 t + s (-6 + t + \epsilon t)}{(-4 + s) (-4 + s + t)},
\end{eqnarray*}
After a change to conformal variables
\begin{eqnarray}
\label{defx}
x(y) = \frac{\sqrt{-s(t)+4}-\sqrt{-s(t)}}{\sqrt{-s(t)+4}+\sqrt{-s(t)}},
\label{our}
\end{eqnarray}
the $a$ has a simple rational denominator:
\begin{eqnarray*}
a
& = & \frac{1}{(1 + x)^2} \frac{1}{ (x + y)}\frac{1}{ (1 + x y)}
 \times \left\{ \cdots \right\}.
\end{eqnarray*}
Such an expression can now be decomposed into terms
with monomial denominators depending on $x$ only,
\begin{eqnarray}
f(0,x) &=& \frac{1}{x}, ~~~~~~~f(1,x) ~=~ \frac{1}{1-x},
\nonumber \\
f(-1,x) &=& \frac{1}{1+x} ,
\label{f}
\end{eqnarray}
and those which depend both on $x$ and $y$:
\begin{equation}
g(-y,x) = \frac{1}{x+y}, \;\;\;g(-1/y,x) = \frac{y}{1 + x y}.
\end{equation}
Factors $b,c,d,e$ in (\ref{box}) follow the same structure.

The monomials become the kernels
for two specific classes of GPLs:
\begin{eqnarray}
\label{gplid}
G(b,\{a\},x) &=&
 \int_0^x dx' g(b,x') G(\{a\},x'),
 \\ \nonumber
 b&=& -y,-1/y .
\label{gpl}
\end{eqnarray}
Analogous relations define the HPLs, now
with the index vector $\{0,1,-1\}$.
If a simpler master was expressed by a structure like (\ref{gplid}), then $m^i$ will be easily determined by iterating the GPLs.

In this way, after performing the $\epsilon$ expansion of the objects in (\ref{box}), the MI {\tt B4l2m} can be solved systematically step by step to the desired order in
$\epsilon$. For more details on the specific example we refer to \cite{Bonciani:2003cj}.
\section{\label{secgpls}Algebra of GPLs}
\subsection{Analytical representation of GPLs}
Up to weight two,
it makes not much effort to find analytical representations for GPLs in terms
of Nielsen polylogarithms.
If we consider only the index vectors $\{-y,-1/y\}$ and $\{0,-1,1\}$,
we have e.g. at weight two 25 GPLs (see Table \ref{table}),
of which we can choose freely 5 as the irreducible
integrals.

We put some results on GPLs up to weight four in file {\tt GPL.m}
at \cite{web-masters:2004np}.
With conventions we follow \cite{Bonciani:2003cj,Czakon:2004tg,Czakon:2004wu,Czakon:2004wm}.
Starting from weight three some of the GPLs in {\tt GPL.m} are
written as numerical integrals, e.g.
\begin{equation}
G[-y,-y,-1,x] = \int_0^x G[-y-1,z]/(y+z) dz.
\end{equation}
Similarly to the case of HPLs discussed in the Introduction, some of
these integrals can be surely solved further to their analytical form.

\begin{table}[h]
\setlength{\tabcolsep}{0.3pc}
\caption{Number of GPLs for the index vectors $\{-y,-1/y\}$ and
  $\{0,-1,1\}$.
The second row counts HPLs with indexes $\{0,-1,1\}$.
}
\vspace*{0.3cm}
\begin{tabular}{|l|l|l|l|}
\hline
      weight 1& weight 2 & weight 3  & weight 4  \\
\hline
      2 GPLs       & 16 GPLs       & 98 GPLs         & 544 GPLs        \\
      3 HPLs       & 9 HPLs       & 27 HPLs         & 81 HPLs        \\
\hline
\end{tabular}
\label{table}
\end{table}
\subsection{Interchange of arguments in GPLs}
Sometimes in the course of solving MIs we want to use the
knowledge of differential equations in a fixed channel, but in both
$s$- and $t$-operators  (or equivalently in $x$ and $y$).
As an example may serve the MI {\tt B5l2m3d3}.
The homogeneous part of its equation, if derived with the
$s$-operator, vanishes, so that its solution is 
a constant.
Then, the solution of the inhomogeneous equation is easy but 
it is not possible to find the constant term (a function $C_y(y)$) in
the usual way by exploiting the knowledge of analyticity. 
However, we can alternatively determine the same function from an
equation derived with the $t$-operator, having now another constant term (a
function $C_x(x)$). 
An easy way to determine $C_y(y)$ is to
compare the functional dependences of both solutions, 
e.g. in the limit $x=1$, and to end up with an unknown true constant $C_x(1)$.
It depends on polynomials in $\zeta_2,
\zeta_3,\ldots$
The $C_x(1)$, in turn,  can be numerically fitted by knowledge of
 {\tt B5l2m3d3} at some 
Euclidean kinematic point using the sector decomposition method \cite{Binoth:2000ps,Binoth:2003ak}.
In this way, {\tt B5l2m3d3} has been determined up to $\epsilon^0$
(yet unpublished). 
Sometimes it may appear from the very beginning more convenient to
solve some differential equations
in the $t$-operator (where e.g. functions $G[-x,...,y]$ appear)
and rewrite this solution later for the $s$-operator where other MIs are
to be solved (and where functions $G[-y,...,x]$ appear).
The interchange of arguments can also be used to find limits of GPLs
when e.g. $x=1$ or $y=1$.

Useful relations for the case of massive Bhabha scattering are tabulated
up to weight four in the file {\tt GPLtransf.m} at \cite{web-masters:2004np}.
All of them are based on the type of identities described in
\cite{Gehrmann:2000zt}.
For instance,  the first item of weight two in {\tt GPLtransf.m},
\begin{eqnarray}
G[-y,-1,x] &=& H[-1, x] H[1, y] \nonumber \\
&-& H[0, x] H[1, y] \nonumber \\
   &+&H[-1, x] H[-x, y] \nonumber \\
&+& H[0, -1, x] \nonumber \\
&+&H[1, 0, y] - G[1, -x, y]
\end{eqnarray}
can be solved using the relation
\begin{eqnarray}
H[\vec{m}(z);y]&=&H[\vec{m}(z=0);y] \nonumber \\
&+&\int_0^z dz' \frac{d}{dz'}
H[\vec{m}(z');y]
\end{eqnarray}
after interchanging the differential and integration operations,
combined with a decomposition into basic polynomials.
\subsection{Conformal transformations}
There are two different definitions in use for
conformal transformations $x$ ($y$) of $s$ ($t$)  in massive QED.
These are (\ref{our}), being used e.g. in \cite{Czakon:2004wm,Bonciani:2003cj},
and
\begin{equation}
x'(y') = \frac{1}{\sqrt{1-4/s(t)}},
\label{smi}
\end{equation}
being used e.g. in \cite{Smirnov:2001cm,Heinrich:2004iq}.
However, $x$ and $x'$ are connected by a relation which is also used for
a transformation of variables in HPLs \cite{Remiddi:1999ew}:
\begin{eqnarray}
x &=&\frac{1-x'}{1+x'},\;\;\; y =\frac{1-y'}{1+y'}.
\label{xp}
\end{eqnarray}
So, if we have to transform MIs defined in terms of $x$
to MIs  defined in terms of $x'$,
we have to know the corresponding relations between
GPLs.\footnote{Results given in 
\cite{Smirnov:2001cm,Heinrich:2004iq} are written directly using
Nielsen Polylogarithms.} 
Here is an example for this:
\begin{eqnarray}
G(-y,0,x) & = & H[ -y,0, 1] + \int_1^x dz \frac{H[0,z]}{z+\frac{1-y'}{1+y'}}
\nonumber \\
& = & H \left[ -\frac{1-y'}{1+y'},0, 1 \right] \nonumber \\
&+& \int_0^{x'} dz [H(1,z)+H(-1,z)] \nonumber \\
& \times & \left[ \frac{1}{1+z}+\frac{y'}{1-z y'} \right] .
\label{ex}
\end{eqnarray}
Interchanging arguments,
\begin{eqnarray}
H\left[-y,0, 1 \right] &=&  - \zeta_2 +H[0,-1,y]-~H[0,0,y], \nonumber
\end{eqnarray}
using conformal transformations for HPLs \cite{Remiddi:1999ew}, and
integrating the expression in (\ref{ex}) we finally get
\begin{eqnarray}
G[-y,0,x] &=&  - \zeta_2 -H[-1,1,y']-H[1,1,y'] \nonumber \\
&-& H[-1,1] ( H[-1,y']+H[1,y']) \nonumber \\
&+& H[0,-1,1]-H[0,0,1] \nonumber \\
& +& H[-1,-1,x']+H[-1,1,x']
\nonumber \\
&+& G[1/y',-1,x']+G[1/y',1,x'].
\nonumber \\
\end{eqnarray}
Let us note that the conformal transformation (\ref{xp}) extends the original set of GPLs by those with additional
arguments $\{+y,+1/y\}$.
Again, results are given in a file {\tt GPLconf.m} at
\cite{web-masters:2004np}.

\bigskip

{\em In conclusion}, we have discussed some basic features of GPLs which are used in an ongoing determination of
fully analytical results in massive Bhabha scattering in QED.
For other interesting properties of GPLs we have to refer to the literature; e.g. analytical continuation of GPLs is
explored in \cite{Gehrmann:2001pz}.
A complete GPL package for general use
in Mathematica will be given elsewhere.

\section*{Acknowledgements}
We would like to thank for partial support
by
Sonderforschungsbereich/Transregio 9--03 of DFG
`Computergest\"utzte Theoretische Teilchenphysik',  by
the Sofja Kovalevskaja Award of the Alexander von Humboldt Foundation
  sponsored by the German Federal Ministry of Education and Research,
and by the Polish State Committee for Scientific Research (KBN)
for the research project in 2004--2005.


\begin{thebibliography}{10}

\bibitem{Kotikov:1991hm}
A.V. Kotikov,
\newblock Phys. Lett. B259 (1991) 314.

\bibitem{Remiddi:1997ny}
E. Remiddi,
\newblock Nuovo Cim. A110 (1997) 1435, hep-th/9711188.

\bibitem{Remiddi:1999ew}
E. Remiddi and J. Vermaseren,
\newblock Int. J. Mod. Phys. A15 (2000) 725, hep-ph/9905237.

\bibitem{Aglietti:2003yc}
U. Aglietti and R. Bonciani,
\newblock Nucl. Phys. B668 (2003) 3, hep-ph/0304028.

\bibitem{Gehrmann:1999as}
T. Gehrmann and E. Remiddi,
\newblock Nucl. Phys. B580 (2000) 485, hep-ph/9912329.

\bibitem{Gehrmann:2000zt}
T. Gehrmann and E. Remiddi,
\newblock Nucl. Phys. B601 (2001) 248, hep-ph/0008287.

\bibitem{Birthwright:2004kk}
T. Birthwright, N. Glover and P. Marquard,
\newblock JHEP 09 (2004) 042, hep-ph/0407343.

\bibitem{Moch:1999eb}
S. Moch and J. Vermaseren,
\newblock Nucl. Phys. B573 (2000) 853, hep-ph/9912355.

\bibitem{Czakon:2004wm}
M. Czakon, J. Gluza and T. Riemann,
\newblock Phys. Rev. D71 (2005) 073009, hep-ph/0412164.

\bibitem{Davydychev:2005nn}
A. Davydychev,
\newblock private communication.

\bibitem{Gehrmann:2001pz}
T. Gehrmann and E. Remiddi,
Comput. Phys. Commun. 141 (2001) 296, hep-ph/0107173.

\bibitem{Maitre:2005uu}
D. Maitre,
\newblock HPL, a Mathematica implementation of the harmonic polylogarithms,
  hep-ph/0507152.

\bibitem{Gehrmann:2001jv}
T. Gehrmann and E. Remiddi,
Comput. Phys. Commun. 144 (2002) 200, hep-ph/0111255.

\bibitem{Bonciani:2003te}
R. Bonciani, P. Mastrolia and E. Remiddi,
\newblock Nucl. Phys. B661 (2003) 289, hep-ph/0301170.

\bibitem{Bonciani:2003cj}
R. Bonciani et~al.,
\newblock Nucl. Phys. B681 (2004) 261, hep-ph/0310333.

\bibitem{Czakon:2004tg}
M. Czakon, J. Gluza and T. Riemann,
\newblock Nucl. Phys. (Proc. Suppl.) B135 (2004) 83, hep-ph/0406203.

\bibitem{Czakon:2004wu}
M. Czakon, J. Gluza and T. Riemann,
\newblock On master integrals for two loop {Bhabha} scattering, hep-ph/0409017.

\bibitem{Aglietti:2004vs}
U. Aglietti,
\newblock The evaluation of loop amplitudes via differential equations,
  hep-ph/0408014.

\bibitem{Fleischer:2003bg}
J. Fleischer, T. Riemann and O. Tarasov,
\newblock Acta Phys. Polon. B34 (2003) 5345,
hep-ph/0508194.

\bibitem{web-masters:2004np}
M. Czakon, J. Gluza and T. Riemann,
\newblock http://www-zeuthen.\linebreak[2]desy.\linebreak[2]de/%
  \linebreak[2]theory/\linebreak[2]research\linebreak[2]/bhabha/.

\bibitem{Binoth:2000ps}
T. Binoth and G. Heinrich,
\newblock Nucl. Phys. B585 (2000) 741, hep-ph/0004013.

\bibitem{Binoth:2003ak}
T. Binoth and G. Heinrich,
\newblock Nucl. Phys. B680 (2004) 375, hep-ph/0305234.

\bibitem{Smirnov:2001cm}
V. Smirnov,
\newblock Phys. Lett. B524 (2002) 129, hep-ph/0111160.

\bibitem{Heinrich:2004iq}
G. Heinrich and V. Smirnov,
\newblock Phys. Lett. B598 (2004) 55, hep-ph/0406053.

\end{thebibliography}

\end{document}